\documentclass[12pt]{article}\usepackage[hyperfootnotes=false]{hyperref}
\usepackage{epsfig}
\usepackage{float}
\usepackage{empheq}
\usepackage{bbold}

\usepackage{caption}

\usepackage{amsmath}
\usepackage{amssymb}
\usepackage{graphicx}
\newlength{\abstractwidth}
\renewcommand{\thefootnote}{\fnsymbol{footnote}}
\renewcommand{\thanks}[1]{\footnote{#1}} 
\newcommand{\starttext}{
\setcounter{footnote}{0}
\renewcommand{\thefootnote}{\arabic{footnote}}}

\newcommand{\be}{\begin{equation}}
\newcommand{\bea}{\begin{eqnarray}}
\newcommand{\eea}{\end{eqnarray}}
\newcommand{\beq}{\begin{equation}}
\newcommand{\ee}{\end{equation}}

\newcommand*\widefbox[1]{\fbox{\hspace{2em}#1\hspace{2em}}}

\def\la{\langle}
\def\ra{\rangle}
\def\simleq{\; \raise0.3ex\hbox{$<$\kern-0.75em
\raise-1.1ex\hbox{$\sim$}}\; }
\def\simgeq{\; \raise0.3ex\hbox{$>$\kern-0.75em
\raise-1.1ex\hbox{$\sim$}}\; }

\def\bi{\begin{itemize}}
\def\ei{\end{itemize}}

\def\CC{{\cal{C}}}

\def\CI{{\cal{I}}}

\def\Tr{\rm Tr \it}
\def\bsub{ \begin{subequations}
\begin{empheq}[box=\widefbox]{align}  }
\def\esub{ \end{empheq}
\end{subequations}}

\def\1{\(  \mathbb{1} \)}
\def\sun{$SU(2^N)$}

  \def\kl{k-local}

  \def\bn{\bigskip \noindent}

\def\aa{\bf{a}\it} 
\def\a{\bf{a}}                
 \def\p{\bf{p}\it}  
 \def\et{Exp-time}   
 \def\eet{Expexp-time}

\makeatletter
\g@addto@macro\normalsize{%
  \setlength\abovedisplayskip{10pt}
  \setlength\belowdisplayskip{20pt}
  \setlength\abovedisplayshortskip{10pt}
  \setlength\belowdisplayshortskip{20pt}
}
\makeatother

\usepackage{color}

\begin{document}


  
\begin{titlepage}

\rightline{}
\bigskip
\bigskip\bigskip\bigskip\bigskip
\bigskip

\centerline{\Large \bf { Black Holes  at  Exp-time }}

\bigskip
\begin{center}
\bf   Leonard Susskind  \rm

\bigskip

SITP, Stanford University, Stanford, CA 94305, USA

Google, Mountain View CA, 94043, USA

\end{center}

\bn

\begin{abstract}

Classical GR governs  the evolution  of black holes for a long time, but at some exponentially   large time it must break down. The breakdown, and what comes after it, is not well understood. In this paper I'll discuss the problem using concepts drawn from complexity geometry. In particular the geometric concept of cut locus  plays a key role.

\end{abstract}

\end{titlepage}

\starttext \baselineskip=17.63pt \setcounter{footnote}{0}

\Large


{\color{blue}\tableofcontents}

\section{Exp-time}\label{sec: Exp-time}

The meaning of the spacetime behind the horizon of a black hole has often been a subject of controversy. One extreme view is that  the world simply ends at the horizon;   everything else  is an unphysical figment. This view leaves unanswered the  question of why classical general relativity predicts a smooth global continuation  past the horizon. Moreover, during the last few years we have gathered evidence that  quantum mechanics allows some access to a black hole's interior. Ignoring the spacetime behind the horizon is not an option and  we have to understand how it emerges holographically.

The question  addressed in this paper is about the  limits of applicability of classical general relativity. There are very good reasons to think that the classical description of a black hole interior must  break down at some exponentially large time (by exponential I 
 mean  in the entropy of the black hole\footnote{A black hole in flat space will long since have evaporated by \et. Throughout this paper the context is black holes in anti de Sitter space, dual to a thermal state of a boundary CFT.  }). If the considerations of this paper are correct the breakdown is marked by a fairly sudden transition.

From now on I will refer to the time scale $t_{exp} \sim e^S$ as \et. The subject of this paper is: What is it that happens at \et; is it sudden or gradual; and what
 exactly replaces the classical description beyond that? 
 To be concrete I'll focus on  the global  volume of the interior of a two-sided black hole in anti-de Sitter space. We'll entertain three  possibilities for how the volume evolves.  The three  agree for early times but  a transition occurs  no later than \et;  beyond that  the three  give different answers. There is no contradiction because the different answers refer to different questions, but only one of the three provides a  geometric picture consistent with the measurable properties of the wormhole.
 
I will use a particular geometric definition of the volume of the black hole interior;  for a one-sided black hole it is the volume of the maximal spatial slice anchored at a given boundary time. Classically it  grows  linearly with time for all time after the black hole has formed. 
For a two-sided black hole the definition is the same except that the maximal slice is anchored at both boundaries, one  at time $t_L$ and the other at time $t_R.$ Classically the volume  grows  proportional to $(t_L + t_R),$ with nothing special happening at \et\footnote{Perhaps a more accurate statement is that classically the entropy of the black hole is infinite, and therefore it never gets to Exp-time.}. The emphasis will mainly be on the two-sided case in which the system is initially in the thermofield-double (TFD) state. The arguments can easily be adapted to the one-sided case.

The identification of volume with complexity is called CV duality. I will assume that it is correct.

\subsection{Clocks}\label{sub: clocks}
There are two kinds of clocks we can appeal to in studying the growth of wormholes. The first are clocks external to the CFT; in other words clocks which are not themselves part of the CFT. They  can  have unlimited accuracy for arbitrary lengths of time. One interpretation  of the volume  is that it is equal to the age of the black hole (as recorded by the boundary clock), multiplied by the area of the horizon. 

A more interesting thing to do is to use the black hole itself as an ``internal" clock. We assume that  the (pure) quantum state of the black hole is characterized by an uncertainty in energy $\Delta E$ equal to the variance of the energy in the thermal state. For a 4-D black hole of radius  $l_{ads}$ the uncertainty in the energy is the Planck mass. More generally it satisfies,
\be 
\Delta E = T\sqrt{S}
\ee
with $S$ being the entropy of the black hole.

As time unfolds the quantum system evolves and passes through a series of mutually 
orthogonal states which can be identified with the states of the growing wormhole. 
The series of orthogonal states can  be used to define a classical internal clock variable.  The time that it takes to pass from one state  to a new orthogonal state is the Aharonov-Anandan time $1/{\Delta E}.$ Let us label the mutually orthogonal states $|V\ra.$ 

The two kinds of clocks, boundary and internal, are expected to agree to high accuracy for  a long period of time but not forever. The number of mutually orthogonal states of the internal clock is bounded by the dimension of the black hole Hilbert space, i.e.,  the exponential of the entropy. Thus by \et, the internal clock will have cycled through all the available states and the subsequent  states of the clock must be  superpositions  of  the earlier  states. 

What happens after $t_{exp}$ depends on the details of the energy spectrum. The average separation of energy levels is 
\be 
\delta E = \Delta E \  e^{-S}.
\ee
If the energy levels were exactly equally spaced the clock would be periodic in time with  period 
$$\frac{e^{S}}{\Delta E}$$
 but that is not what is expected for black holes. Black holes are chaotic systems with random-matrix energy spectrum. For chaotic systems the state after  \et  \ will become  a superposition of  the linearly independent earlier clock states with exponentially small amplitudes,
\be 
|\rm clock \it \ra_{t>t_{exp}} = \sum_{i=1}^{e^S} f(V_i, t)|V_i\ra  \ \ \ \ \ \ \ \ |f(V,t)|^2 \sim e^{-S}
\ee

The framework for what follows is the two-sided black hole,  dual to two entangled copies of a holographic CFT. The clock variable $V$ will be the volume of the Einstein-Rosen bridge (or wormhole) connecting the two black holes. Different values of $V$ correspond to different classical wormhole geometries of different volume.

The fact that the clock state for $t>t_{exp} $ becomes a superposition of states with different volume indicates a massive breakdown of classical GR after Exp-time. It suggests that there will be  no concept of a single classical geometry, but only a quantum superposition of many  macroscopically different geometries \footnote{The phenomenon of ``running out of states" occurs in many contexts including the theory of giant gravitons   \cite{McGreevy:2000cw} as well as the theory of Euclidean wormholes    \cite{Marolf:2020xie}. I thank Steve Shenker for pointing this out.   The fact that wormhole growth is ultimately bounded by  the dimensionality of the Hilbert space, and that very old wormholes must be linear superpositions of shorter wormholes has been recognized and discussed by P. Saad, S. Shenker, and D. Stanford (unpublished). }.

From a boundary point of view the superposition of geometries is a technically correct way to describe the quantum state of an exponentially old wormhole, but for reasons  that I'll explain, this   does not preclude there being a single classical geometry of the interior.

\subsection{Complexity}\label{sub: complexity}
The complexity-volume  correspondence\footnote{One could equally well use the complexity-action correspondence.} is a duality between the growth of Einstein-Rosen bridges (a.k.a. wormholes) and the time evolution of quantum computational complexity\footnote{From now on just complexity.}. We have little or no knowledge about how wormholes evolve over very long periods of time, but we do know something about how complexity evolves. One can hope to leverage this knowledge into a theory of wormhole evolution.

 We will consider two-sided black holes and model them as maximally entangled states of $2N$ qubits. Such a state may be written in the form,
\be 
|\Psi\ra = \sum U_{ij}|i,\bar{j}\ra
\ee
where the index $i$ labels a basis for the right system and $j$ labels the time-reversed basis  for the left system. The state-complexity of $|\Psi\ra$ is equivalent to the unitary operator complexity of $\sum U_{ij}| i \ra \la j |.$  Thus from a mathematical standpoint   the evolution of the state complexity of the $2N$-qubit state $|\Psi\ra$ may be replaced by  the evolution of the complexity  of $N$-qubit unitary operators.

\begin{figure}[H]
\begin{center}
\includegraphics[scale=.3]{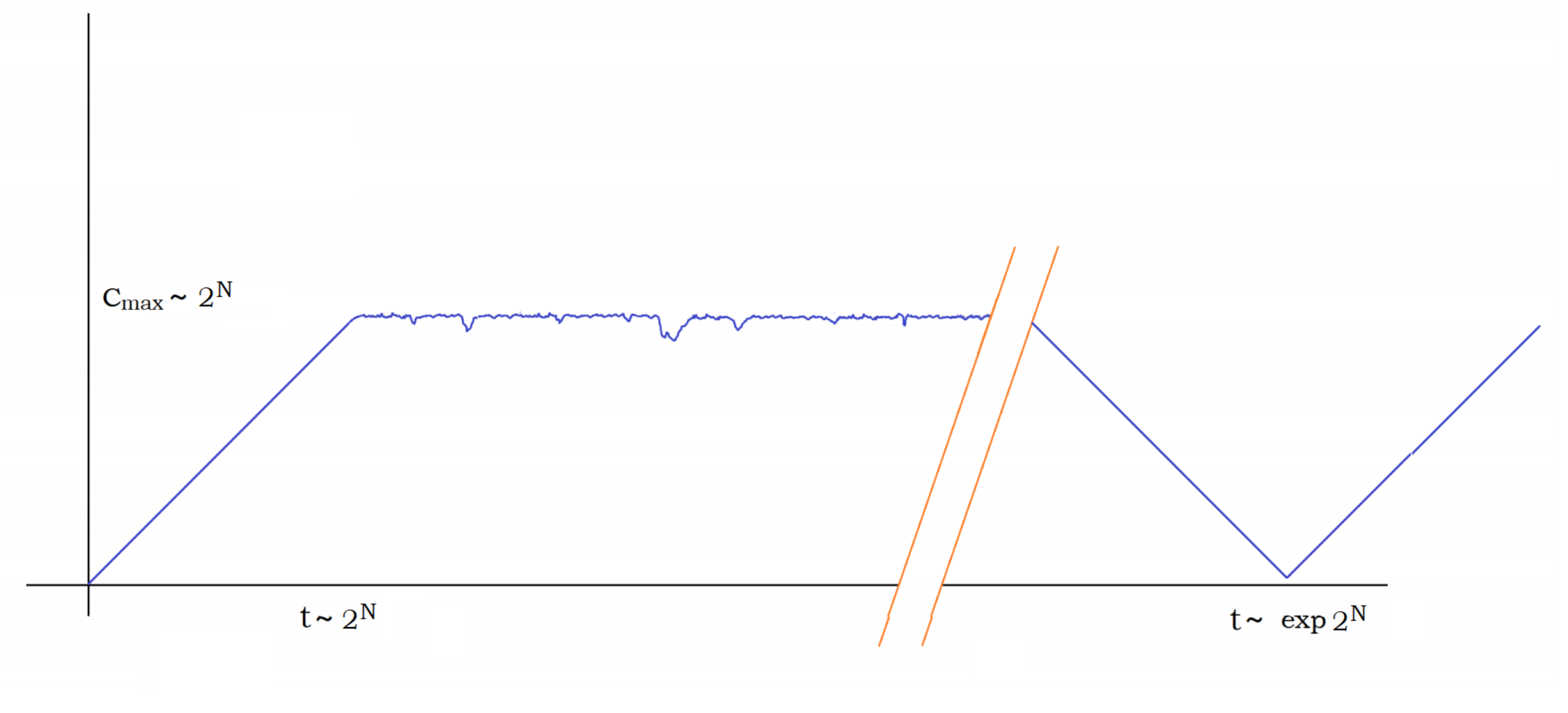}
\caption{Time-dependence of unitary operator complexity  for a chaotic system.}
\label{C-evolution}
\end{center}
\end{figure}

The conjectured  curve for the evolution of complexity of a chaotic system  is well known   \cite{Susskind:2018pmk} (See figure \ref{C-evolution}). 
\bn
The curve represents unitary operator complexity $\CC(t)$ as it evolves according to some \kl \ Hamiltonian. It also represents the growth of complexity for a maximally entangled 2-sided black hole. The complexity (according to the conjecture) increases linearly for an exponential time. In analogy with the behavior of the spectral form factor   \cite{Cotler:2016fpe}\cite{Gharibyan:2018jrp}\cite{Saad:2018bqo} I'll call this linear growth region the \it complexity ramp\rm\footnote{The similarity of the complexity curve and the spectral form factor curve is probably coincidental and does not reflect any similarity of the physics.
The growth of complexity has to do with the growing separation  of the initial and evolving states in the complexity metric. The ramp in the spectral form factor is connected with a decrease of the distance between the two states in the inner product metric. The arguments of section \ref{sec: Cgeometry} suggest a very sharp transition between complexity ramp and plateau. By contrast the spectral form factor transition is probably much broader. }.

\begin{figure}[H]
\begin{center}
\includegraphics[scale=.3]{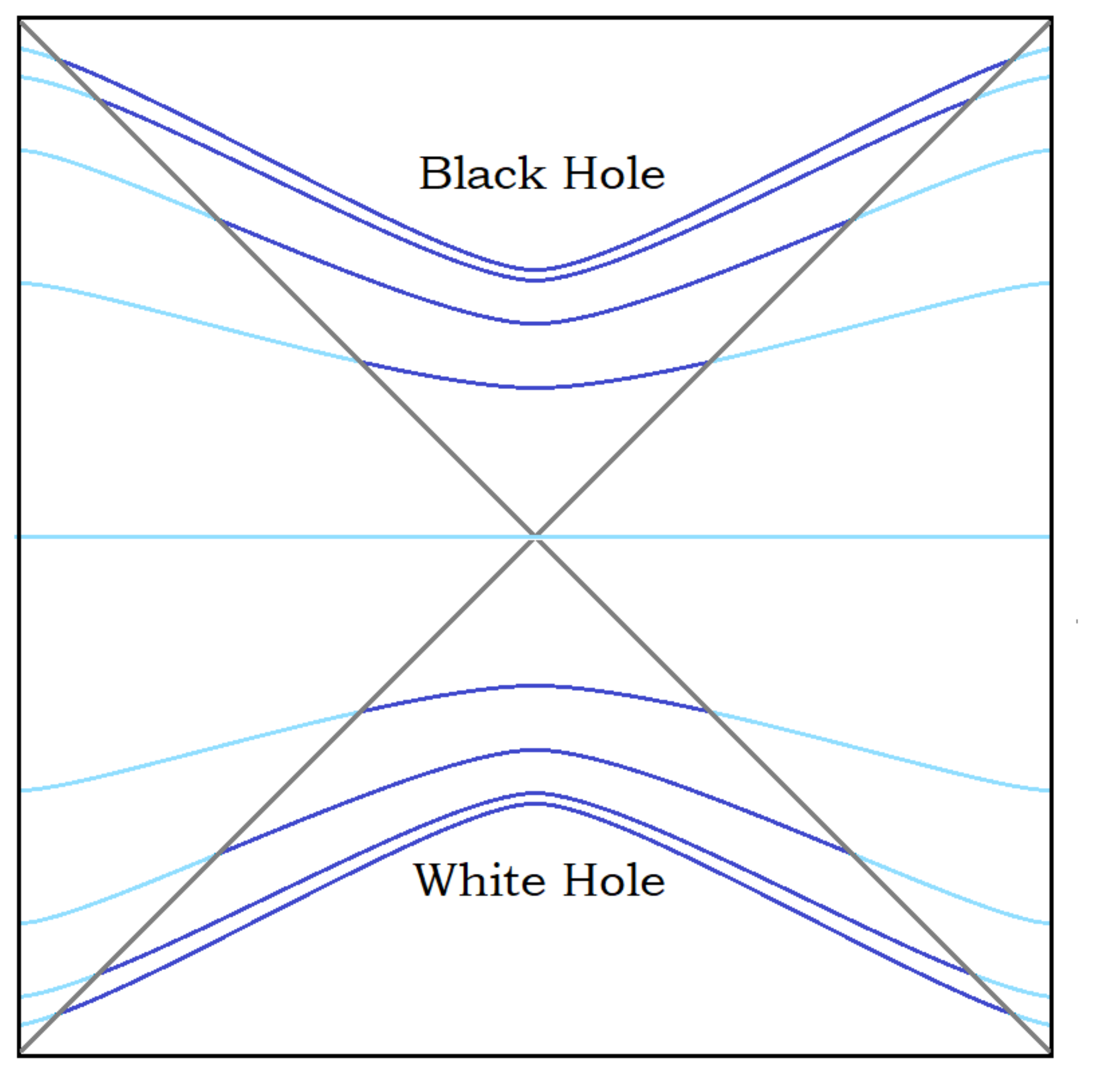}
\caption{Penrose diagram for two-sided eternal black hole. The diagram has been foliated by
maximum volume slices.}
\label{Whiteblack}
\end{center}
\end{figure}

The assumed linear  complexity growth  parallels the classical  wormhole growth illustrated in the upper half of the Penrose diagram in figure  \ref{Whiteblack}.
But unitary operator complexity is bounded, so the complexity curve must eventually stop growing.
How large the complexity becomes before saturating is subtle. The maximum complexity of any unitary operator is $\sim 4^N$ which also happens to be the dimension of the space \sun. One might expect that in a time of order $4^N$ the system will reach maximum complexity.

The  volume of \sun \ (measured in $\epsilon	$-balls \cite{Susskind:2018pmk})  is doubly exponential , 
\be 
\rm Vol\it_{SU(2^N)} \sim e^{4^N}
\ee
 implying that the time to reach the neighborhood of every point  is  $e^{4^N}.$
Indeed both of these things  would be  correct if the system evolved by random circuit, or Brownian circuit, dynamics.

However, evolving by a time-independent Hamiltonian puts extra restrictions on what operators can be reached.  $U(t)$ has the form,
\be 
U(t) = \sum_n e^{i\theta_n} |E_n\ra\la E_n|
\ee
where the phases $ e^{i\theta_n}$ are given by
\be 
e^{i\theta_n} = e^{-iE_n t}
\ee
It follows that the point $U(t)$ is restricted to lie on a $2^N$-dimensional torus embedded in the $4^N$-dimensional \sun. If the $E_n$ are incommensurable, i.e., their ratios are irrational, then the motion of $U(t)$ will ergodically fill the torus in a time $e^{2^N}.$ This suggests that the largest complexity that can be reached by time-independent Hamiltonian evolution is $2^N.$ It's for this reason that I defined $t_{exp} \sim 2^N.$

At this point the complexity ramp gives way to the \it complexity plateau.\rm \
On the complexity plateau the system is in \it complexity equilibrium \rm with complexity 
$\CC= 2^N.$   It stays that way for a doubly exponential recurrence time that I'll denote  ``Expexp-time"  $ t_{expexp} \sim e^{ 2^N}.$  In section \ref{sec: Cgeometry} we will see evidence that  the crossover from complexity-ramp to complexity-plateau is sharp.

From the bulk point of view, what if anything happens at  Exp-time? The complexity-volume correspondence (CV), if one believes it for such long times,  implies that the wormhole volume also reaches a plateau and stops growing.
This raises the question: What is the bulk mechanism that accounts for this breakdown of classical GR? 

At best I will only give  a partial answer in this paper.

\section{Quantum Recurrences and the Full Penrose Diagram}

If one waits  Expexp-time $\sim e^{2^N}$ one will eventually see a quantum recurrence whose duration is
\be 
\Delta t= 2 t_{exp},
\ee
the factor $2$ representing the times for the complexity to decrease and then increase.
  This is illustrated by the V-shaped portion of the curve in figure \ref{C-evolution}  which is centered  at a time $ t\sim e^{2^N}.$ The quantum recurrence is of course an extraordinarily rare event: partial recurrences are vastly more likely than a full recurrence in which the complexity returns all the way to its initial value. 
But conditioning on the assumption that the complexity does return to the initial value, the most likely way for it to do so is by the V-shaped portion of the curve. The quantum recurrence is a  version of an extreme Boltzmann fluctuation, but involving complexity rather than entropy \cite{Brown:2017jil}.

Quantum recurrences provide a new perspective on the full Penrose diagram in figure \ref{Whiteblack}.  The white hole portion of the diagram is often considered to be unphysical, to be replaced by some process that creates the TFD state at $t=0.$ However, it is clear from  figure  \ref{Whiteblack} that the volume of the wormhole tracks the same history as the quantum recurrence.  Past and future infinity  represent complexity equilibrium and the rest of the diagram represents the complexity version of the extreme Boltzmann fluctuation. To my knowledge this interpretation of the  full diagram in terms of a complexity recurrence  has not previously been given.

Note that the duration of the fluctuation is 
 a mere Exp-time embedded in a much longer Expexp-time. It is during these extremely sparse periods that classical GR  describes the geometry behind the horizon.

 \section{The Geometry of Complexity}\label{sec: Cgeometry}

We now come to the main subject of this paper: the implications 
of  complexity geometry for the evolution of chaotic systems over exponential and doubly exponential times. In particular we will be interested in what it can teach us about the transition from complexity ramp to complexity plateau, the fluctuations on the plateau, and quantum recurrences. Granting  CV duality we will also be learning lessons for the evolution of wormhole geometry.
I will assume the reader has some familiarity with complexity geometry both in the original form \cite{Nielsen:2007aaa},  and as applied to black holes in \cite{Brown:2016wib}\cite{Brown:2017jil}. 

A geometric ingredient that will play a key role in understanding the complexity ramp-plateau transition is the concept of \it cut locus \rm which I will now review.

\subsection{Cut Points and Cut Locus}
Consider a compact Riemannian geometry\footnote{The ideas of cut points and cut loci are also applicable to many other metric geometries such as Finsler and sub-Riemannian geometries. }.
Define a distance function for pairs of points $\p, \a,$
\be 
\rm distance \it = L(\p, \a)
\ee
The distance function is the length of the \it shortest \rm geodesic connecting the two points. The distance function has a maximum value which is called the diameter of the space.
We  let $t$ be a path parameter along a geodesic which measures length.

\begin{figure}[H]
\begin{center}
\includegraphics[scale=.5]{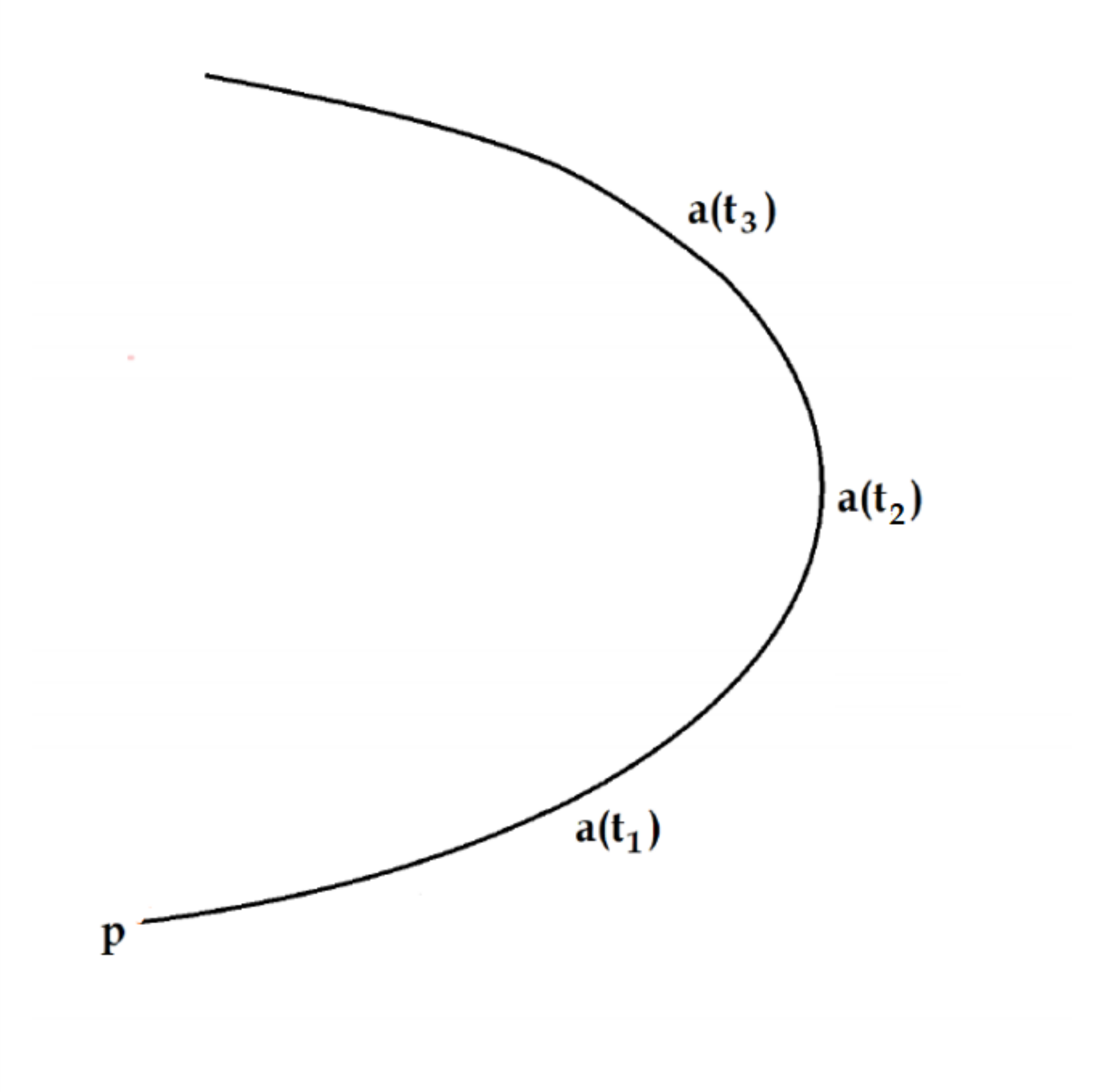}
\caption{A geodesic   $\aa(t)$ originating at $\bf p$ parameterized by $t.$ }
\label{nocut}
\end{center}
\end{figure}

To define cut points and the cut locus
 of a point $\bf p,$ we consider all the geodesics emanating from that point. Let's pick one and call it $\a\it (t)$. This is illustrated in figure \ref{nocut}

\begin{figure}[H]
\begin{center}
\includegraphics[scale=.4]{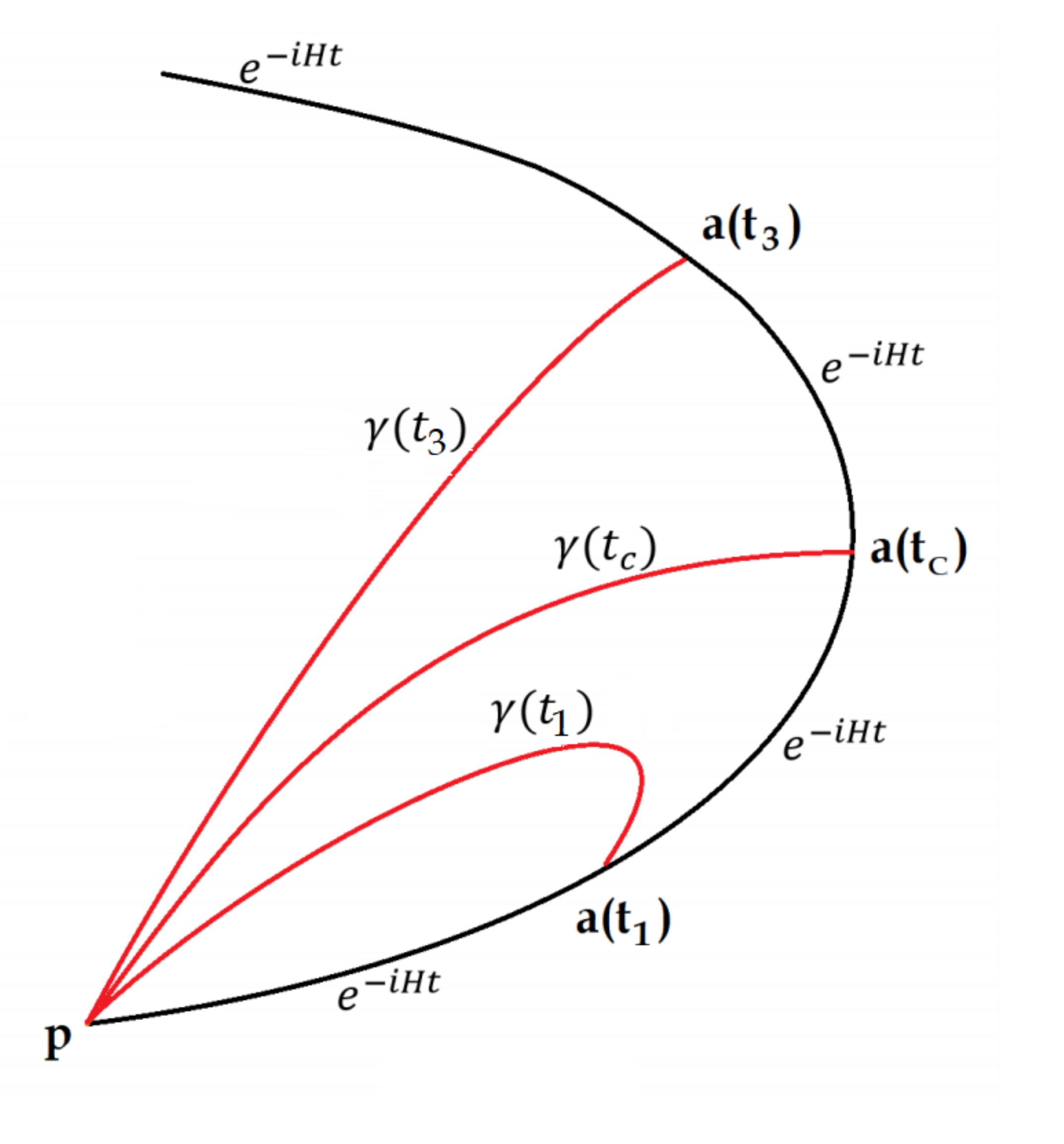}
\caption{For $t<t_c$ the shortest geodesic connecting $p$ with $\aa(t)$ is shown in black. After the geodesic passes the cut locus at $t_c$ the shortest geodesic at any given $t$ is the red curve $\gamma(t).$}
\label{cut}
\end{center}
\end{figure}

 For small enough but finite $t$ it is certain that $\a\it (t)$ is the shortest geodesic connecting $\bf p$ to the point $\a\it (t).$ But at some point---the cut-point  labeled $t_c$ in figure \ref{cut}---a second geodesic, $\gamma(t_c)$, of the same length as $\a\it (t_c),$   may intersect $\a\it (t)$. Past $t_c$ the family of red geodesics, $\gamma(t),$ replaces  $\a\it (t)$ as  the minimal  geodesics defining the distance function $ L(t) \equiv \  L \Big( \p, \aa(t)\Big)$.
 It is important to understand that  $\a\it (t)$ is a single geodesic,  but $\gamma(t) $ represents a family of geodesics connecting $\p$ to the ``moving" point $\aa(t)$. After the cut point no single geodesic gives the length function $L(t).$

Note that at the cut point the length function  $L \Big( \p, \aa(t)\Big)  $  
 is a continuous function of $t$  but the first derivative is not. 

\bn

The cut locus of the point  $\bf p$ is the set of cut-points that are obtained by replacing 
$\a\it (t)$ by the set of all geodesics through $\bf p.$  For homogeneous geometries the
 structure of the cut locus is independent of $\bf p.$ That is the case for complexity
   geometries which inherit the homogeneity from the group structure of \sun.  	It is therefore
     sufficient to understand the cut locus of the identity.
     
\begin{figure}[H]
\begin{center}
\includegraphics[scale=.3]{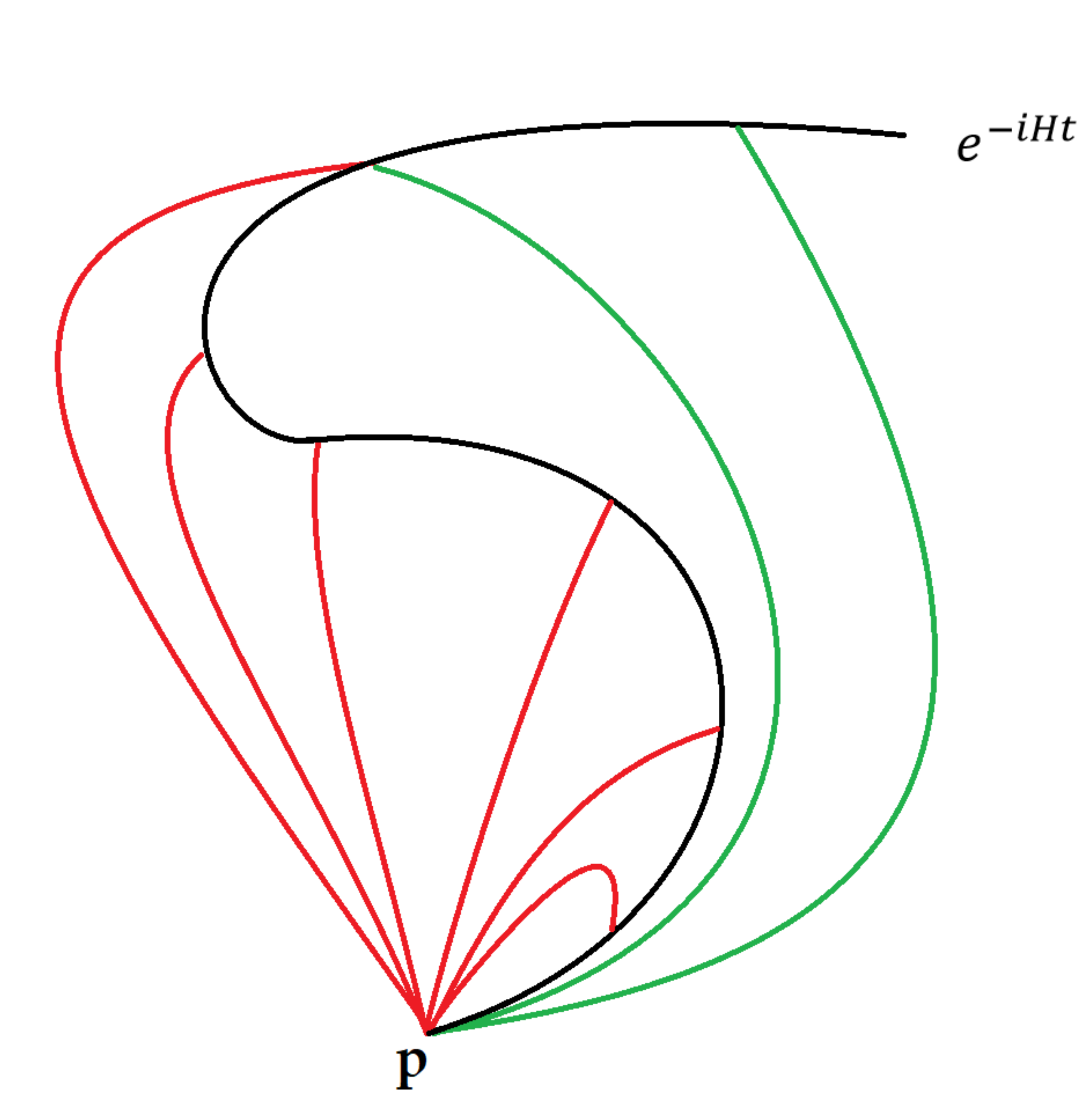}
\caption{As the black geodesic continues on its way, the shortest geodesic connecting $p$ with $\aa(t)$ may pass through a number of cut points at which new families of geodesics come into play.}
\label{cutcut}
\end{center}
\end{figure}

     Things get more complicated if we continue $\a\it(t)$ further as in figure \ref{cutcut}. At some new point a second cut may occur in which another geodesic, shown in green,  becomes shorter than the corresponding member of the red family. Again there is no discontinuity in the length, but there is  a discontinuity 
in the derivative of the length. From that point forward the green family defines the shortest geodesic, until the next cut is reached.

If we define the parameter $t$  along the original geodesic---black in figures \ref{cut} and \ref{cutcut}---then the shortest distance from  $\bf p$ to the moving point   $\a\it(t)$ is $L\Big(\p,\a(t) \Big).$ In general it has a rather 
 complicated behavior. Before the cut point  it is linear because the original geodesic is the shortest. When the cut locus is reached the distance function will be continuous but the first derivative will  not be. The length function may not even be monotonic. As we move forward on $\a\it(t)$ there may be a series of cuts in which the derivative $dL/dt$ jumps. 

\subsection{Complexity Geometry} 

Complexity geometry was introduced by Nielsen and collaborators  \cite{Nielsen:2007aaa} and applied to black holes in \cite{Brown:2016wib}\cite{Brown:2017jil}. 
   A complexity geometry is a right-invariant Riemannian  geometry on the group manifold \sun---the manifold  representing the space of unitary operators acting on a system of $N$ qubits. The metric of any right-invariant Riemannian geometry has the form \cite{Brown:2017jil},
\be 
dl^2=d\Omega_I  \ \CI_{IJ} \  d\Omega_J,
\label{right invariant}
\ee
where
\be 
d\Omega_I =i \Tr  d U^{\dag} \sigma_I U.
\label{dOmega}
\ee
where the notations are those of \cite{Brown:2017jil}. I list them here:
\bi 
\item $\Tr$ indicates normalized trace defined so that the  $\Tr$ of the identity element is $1.$
\item The subscripts $I, J$ label the generators of \sun \ in the Pauli basis. 
\item $\sigma_I$ denotes an element of the $(4^N-1)$-dimensional Pauli algebra. Each $\sigma_I$ is a monomial  composed out of the $3N$ Pauli operators describing the $N$ qubits.  No qubit appears more than once in the monomial.   The weight\footnote{The weight of a monomial is the number of qubits that appear in the monomial. It ranges from $1$ for single qubit operators to $N$ for a product involving all the qubits.} of the Pauli operator $\sigma_I$ is called $w_I.$
\item  The matrix $\CI_{IJ}$ is  symmetric. It is assumed to be diagonal and a function only of the weight.
\be 
\CI_{IJ} = \delta_{IJ} \CI(w_I)   
\label{inner}
\ee
The usual inner product metric\footnote{The inner product metric on \sun \ is defined by a distance function given by $d(U,V) = \arccos|\Tr U^{\dag}V|$ where $\Tr$ means normalized trace.} on \sun \ is a special case of \ref{right invariant} in which,
\be 
\CI_{IJ} = \delta_{IJ}. 
\ee

The function $\CI(w_I)$ is called the penalty factor or cost factor. It represents the complexity cost of moving in the direction $I.$ It is assumed to grow rapidly 
with the weight $w_I.$

\item
In the original version of complexity geometry   \cite{Nielsen:2007aaa}
 The function $\CI(w_I)  $ was
taken to be unity for \kl \ (easy) directions for some fixed $k$, and either infinite or exponentially large (order $4^N$) in all other (hard) directions. There are many reasons to think that this penalty schedule is too severe, some of which were described in \cite{Brown:2017jil}.  Among them     three facts stand out: 
\bi 
\item The geometry based on the original penalty schedule is extremely singular with either infinite or exponentially large  sectional curvatures. 
\item The geometry  is fractal-like with anomalous Hausdorff dimension.
\item In some directions the cut locus of a point $p$ comes very close to that point, but this reflects the rough texture of geometry on small scales, and not any interesting physical phenomena   \cite{Brown:2019whu}.
\ei

The singular features of the original geometry are not inevitable for  a good description of complexity.
In   \cite{Brown:2017jil} arguments were given for a smoother dependence of $\CI(w_I),$ namely, an exponential growth, 
\be 
\CI(w) = e^{\alpha w}
\label{exp-penalty}
\ee
with $\alpha$ a fixed $N$-independent constant of order $1$.
\ei
The important features of this geometry---some proved, one conjectured---include the following   \cite{Brown:2017jil}\cite{Brown:2019whu}\cite{Freedman}.

  \begin{enumerate}
  \item The geometry is homogeneous (everywhere the same). This is insured  by the group properties of \sun.  \  (Proved---trivially)
  \item The diameter of the geometry (the maximum distance between points) is exponential in the number of qubits.   \  (Proved not so trivially in \cite{Freedman})  \rm
  \item The volume of the geometry is exponential in the diameter. In other words it is doubly exponential in the number of qubits.   \  (Proved---easily \cite{Susskind:2018pmk})
  \item Generically, sectional curvatures are negative for sections determined by pairs of low-weight directions. The curvature is of order $1/N$  which is what is need to describe the switchback effect \cite{Brown:2016wib}.     \  (Proved by calculation \cite{Brown:2017jil})
\item The previous four items are all rigorously established for the geometry with exponentially growing penalty factors. The next is a conjecture which is at the heart of the arguments that follow but is unproved.

\bn
\bf Conjecture \rm
\bn 

The cut locus  in generic directions\footnote{There are non-generic directions in which the cut locus is much closer to the identity. For example directions defined by monomials of the Pauli operators are periodic and the cut locus is small for low weight directions. In most directions the geodesics are not closed curves and are infinite.} 
for the geometry defined by the penalty factors \ref{exp-penalty} 
is at the maximum possible distance, i.e, the diameter $\sim 2^N.$ 
  \end{enumerate}
  
  \bn
  
  The status of this conjecture is unclear at the moment. 
  It is largely based on the arguments of \cite{Susskind:2018pmk} where I explained the relation between quantum circuits and expander graphs. Proving the conjecture is probably a very difficult mathematical problem but one can hope that in time more evidence for it will  accumulate. But even if it proves to be false, the connection between cut loci and the evolution of complexity remain intact. What is at stake is not that connection but rather the detailed structure of the curve in figure \ref{C-evolution}. A series of cuts at smaller distance from the identity might spread out  the conjectured sharp transition between the complexity ramp and the plateau.

  \subsection{Analogy with Expander Graphs}\label{subsec: expander}
  The properties that I just described are in close analogy with the properties of regular expander graphs \cite{Lin:2018cbk}\cite{Susskind:2018pmk}. Let me spell out the analogy by listing the properties of homogeneous expander graphs. The list parallels the list of properties 1-5 in the previous subsection.
  \begin{enumerate}
  \item Homogeneous expander graphs are everywhere the same. In other words they look the same when viewed from any vertex. In particular the degree is the same at every vertex.
  \item The diameter (maximum graph distance between any two points) is finite.
  \item The concept of volume is played by the total number of vertices in the graph. The volume is exponential in the diameter.
  \item As viewed from any vertex the graph is locally tree-like. As one moves outward from a vertex the number of vertices at a given distance grows exponentially. This is the analog of negative sectional curvatures for the complexity geometry case.
  \item  Closed loops parametrically smaller than the diameter of the graph are absent or very rare.
  \end{enumerate}
  
  On the last point, let's consider starting at a vertex and working outward along two different branches of the tree. Eventually we may encounter a collision at which the branches reach the same vertex. This will form a loop. The principle of \it no small closed loops \rm translates to the statement that no collisions take place until distances of order the diameter are reached. This is very similar to the idea that the cut locus of a geometry in most directions is at distance of order the diameter.
  
 Coming back to complexity geometry,  items 1---5 may be summarized as:
 
\bn  
  \it Complexity geometry is a right-invariant expander geometry on the group \sun. \rm
  
  \bn

 Constructing strictly homogeneous expander graphs is difficult but there is a weaker condition that is not so hard to implement: 
  Statistically homogeneous expander graphs  are on-the-average homogeneous and 
  look statistically the same from any vertex. For example the degree (number of edges) at each vertex, or some other local quantity  may fluctuate about a mean with a small variance.

\begin{figure}[H]
\begin{center}
\includegraphics[scale=.25]{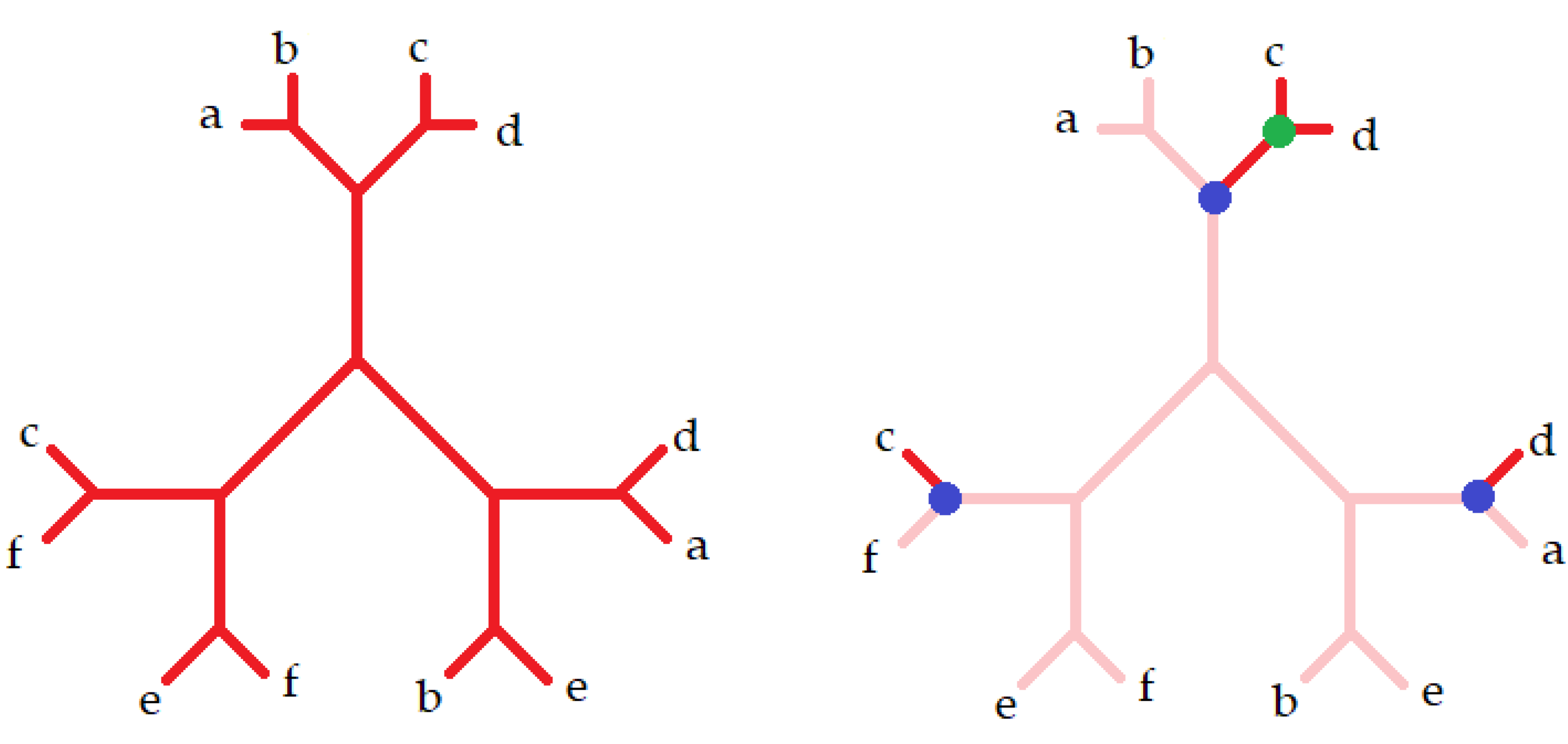}
\caption{Construction of an expander graph. The leaves of a tree of depth $3$ are identified as in the left panel. In the right panel
a vertex in green is shown along with its three nearest neighbors in blue. The edges connecting them are shown in darker color.}
\label{TREE}
\end{center}
\end{figure}

  Statistically homogeneous expander graphs can be built by starting at a point and constructing a tree of a given large depth.   If one then randomly identifies the leaves as in figure \ref{TREE} the result will be a statistically homogeneous expander graph. Of course this really applies to very large graphs with many vertices and edges.

  \subsection{Geodesics in Complexity Geometry}\label{subsec: Geodesics in Complexity Geometry}
  
Given a sum of generators $H=\sum_{I} h_I\sigma_I$ a curve can be swept out by exponentiation, $$U(t) = e^{-iHt}.$$     If the geometry is defined by the usual inner product metric then the curve will be a geodesic for any $H.$ This is not the case for more general metrics of the form described in equations  \ref{right invariant}, \ref{dOmega}, and \ref{inner}.
The necessary and sufficient conditions for $e^{-iHt}$ to be a geodesic of the complexity metric is that $h_I$ be an eigenvector of the matrix $\CI_{IJ}.$ This means that all the $\sigma_I$ in the sum must have the same weight. I will assume  the real Hamiltonian describing a black hole  is of that form. Thus the history traced out by the actual Hamiltonian evolution of a black hole will be a geodesic. But we will need to keep in mind that most Hamiltonians are of mixed weight and do not generate geodesics---and that most geodesics are not made by exponentiating time-independent Hamiltonians.

Any curve can of course be generated from a time-dependent Hamiltonian by time-ordered exponentiation,
\be 
U(t) = T e^{-i\int_0^t dt' H(t').}
\ee
Geodesics generated in this way are not homogeneous along their length. 

Returning to figure \ref{cutcut}, let's assume that the original geodesic $\aa(t)$ was generated 
by a time-independent Hamiltonian of fixed weight.  Once we have passed the cut locus the shortest geodesic is no longer $\aa(t),$ but rather a member of the set of red geodesics $\gamma(t).$ In general these geodesics will lie along directions of mixed weight, and as a result they will be generated by time-dependent Hamiltonians. They will not be homogeneous along their length.

\section{Simple Model of Complexity geometry}\label{sec: simple model}

The expander property of complexity geometry can be illustrated in a  toy model described in \cite{Brown:2016wib}.
Although highly simplified the model has all the features 1---5 explained  in section \ref{subsec: expander}. Most of  what follows applies  to any version of complexity geometry that has the expander property.

\begin{figure}[H]
\begin{center}
\includegraphics[scale=.3]{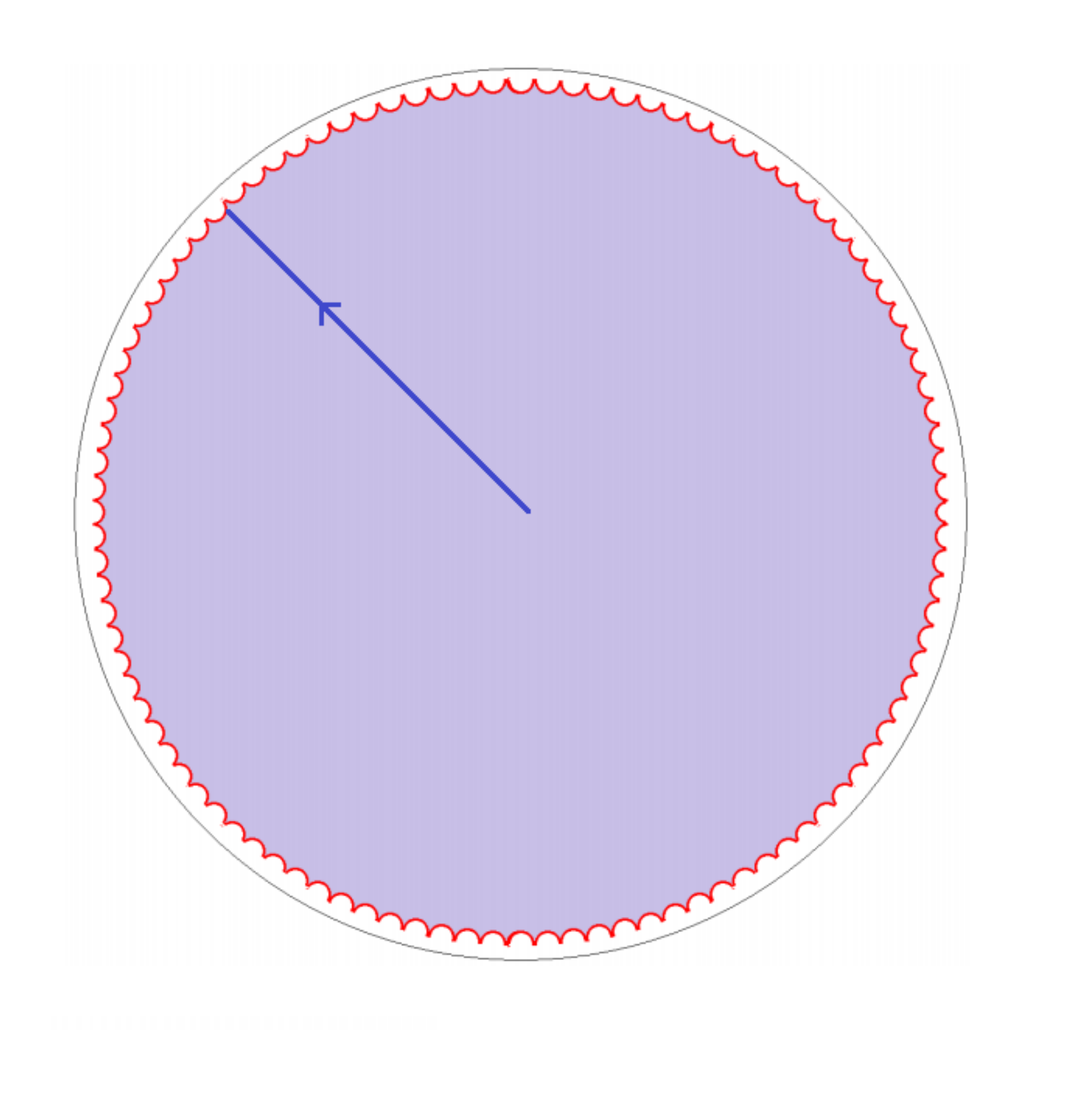}
\caption{Hyperbolic disc with an inscribed regular polygon.  The blue line is a geodesic beginning at the identity operator.}
\label{disc1}
\end{center}
\end{figure}

The construction of the model  is entirely analogous to the construction of statistically homogeneous expander graphs. Instead of a tree with leaves we begin with a region of the hyperbolic disc bounded by  a regular polygon of  $e^{2^N}$ sides. The sides  play the role of the leaves of the tree.
 This is shown  in figure \ref{disc1}. 

 The next step is to randomly identify the sides (there is a weak constraint that is necessary to avoid a conical singularity). The resulting space is a Riemann surface having the following properties:
\begin{enumerate}
\item The total area of the geometry is is $e^{2^N}$ which is of the same order as the volume of \sun. The area  is the analog of the number of vertices in the graph analogy.
\item The genus of the Riemann surface is $e^{2^N}$.
\item The diameter (maximum distance between points) is $2^N,$ the same as the maximum complexity generated by a time-independent Hamiltonian for $N$ qubits. 
\end{enumerate}

The points of the geometry, labeled $\a,$ schematically represent unitary operators, the center of the disc representing the identity operator. The geodesic distance from the center defines the complexity of a point . 

 In figure \ref{disc1}   the blue directed line  represents a geodesic generated by a \kl \ Hamiltionian $H$ of definite weight, i.e., $\aa(t)=e^{-iHt}$. (In the real complexity geometry different Hamiltonians determine different directions away from the origin.)   The trajectory begins at the identity operator,  $\aa(0)=I,$  and moves outward toward greater  complexity. 

\begin{figure}[H]
\begin{center}
\includegraphics[scale=.5]{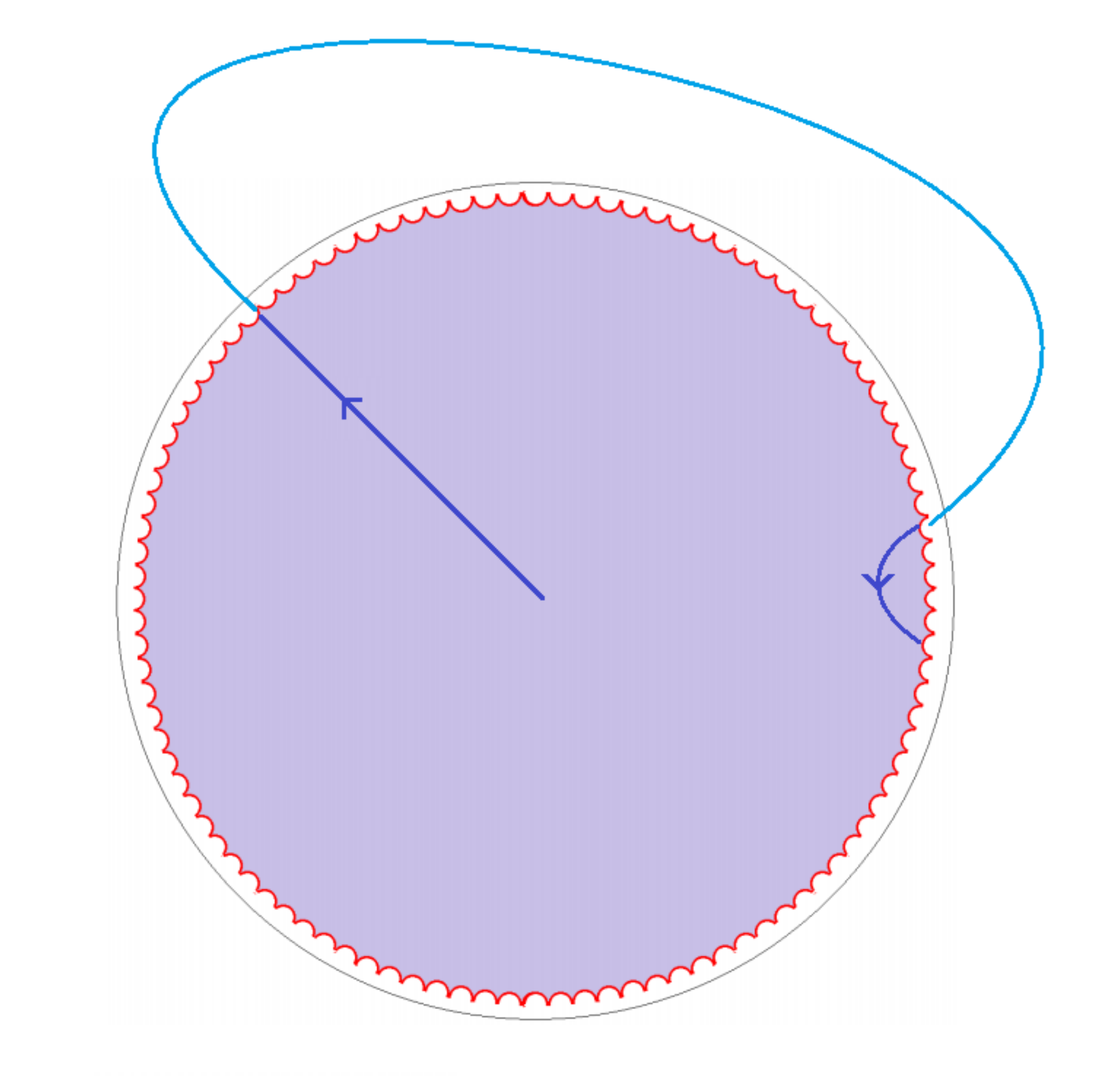}
\caption{The geodesic $e^{-iHt}$  represented  in dark blue passes through the polygon and reappears at a point determined by the identification rule. The light blue curve is just to guide the eye.}
\label{disc2}
\end{center}
\end{figure}

The geodesic $\aa(t)$ does not hit a cut locus before reaching the polygon.
 When it does reach the polygon it exits and reappears, moving inward at another location. This is shown in figure \ref{disc2}.

\begin{figure}[H]
\begin{center}
\includegraphics[scale=.5]{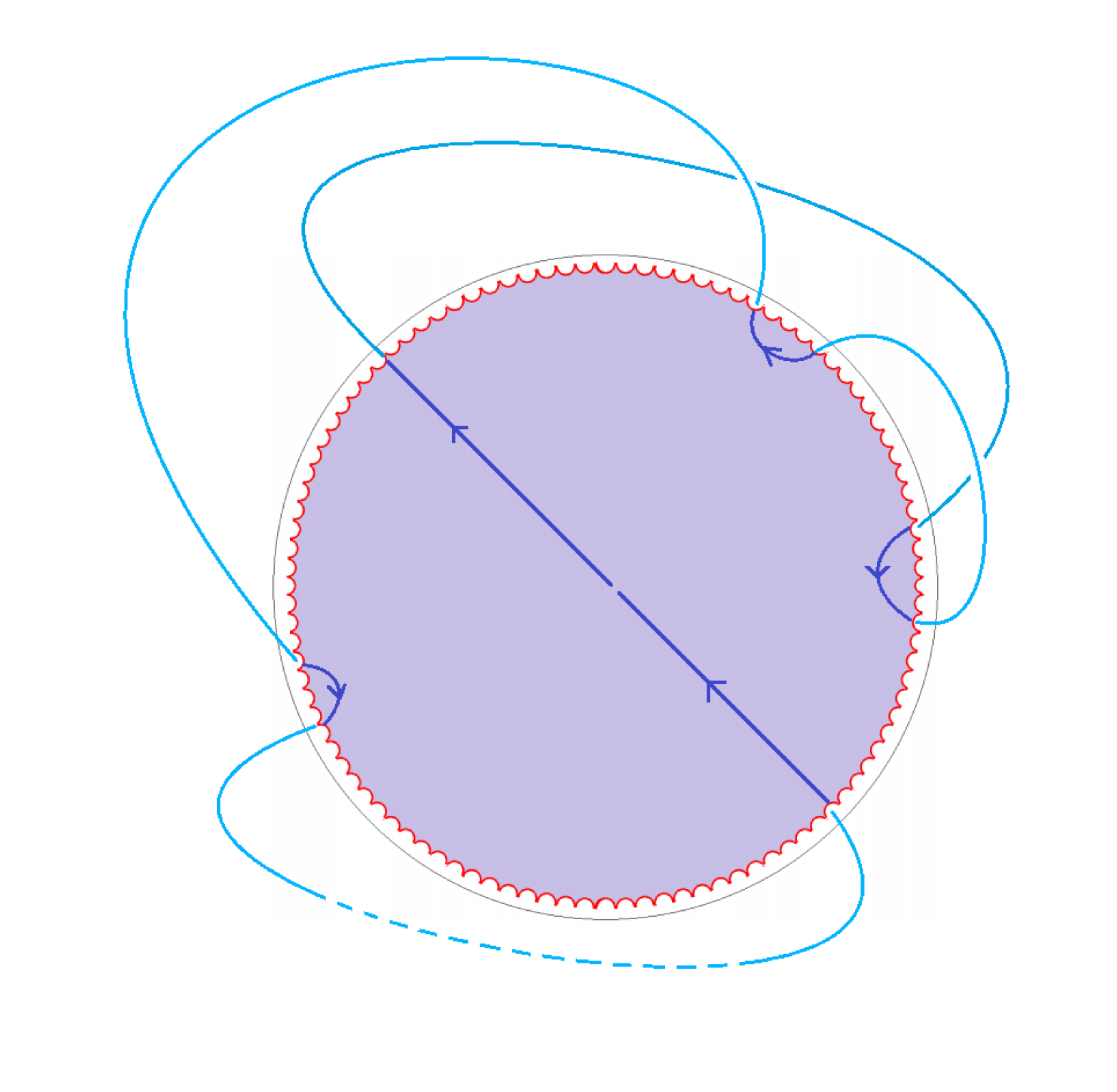}
\caption{After many ``jumps" the geodesic will eventually execute a quantum recurrence. }
\label{disc4}
\end{center}
\end{figure}

\bn
The precise angle at which the ingoing line re-enters the geometry   depends on exactly where the outgoing line hit the side of the polygon.
Unless it is finely tuned to one part in $e^{2^N}$ the trajectory 
 will quickly  turn around and again hit the polygon after a short interval. The interpretation is that once the system reaches maximal complexity it will bounce around among the exponentially complex states for a very long (doubly exponential in $N$) time. 
This will  repeat itself until by accident a recurrence occurs as in figure \ref{disc4}

\bn
This model reproduces all of the features of  the complexity curve shown in figure \ref{C-evolution}: along the initial radial line the complexity increases linearly; once $\a(t)$ reaches the polygon it exits and reenters the geometry at a new point, and with overwhelming probability it quickly turns around and exists a second time. This entering and exiting produces the jagged plateau in figure \ref{C-evolution};  and very rarely, on an \eet \ time scale a recurrence occurs.

\begin{figure}[H]
\begin{center}
\includegraphics[scale=.4]{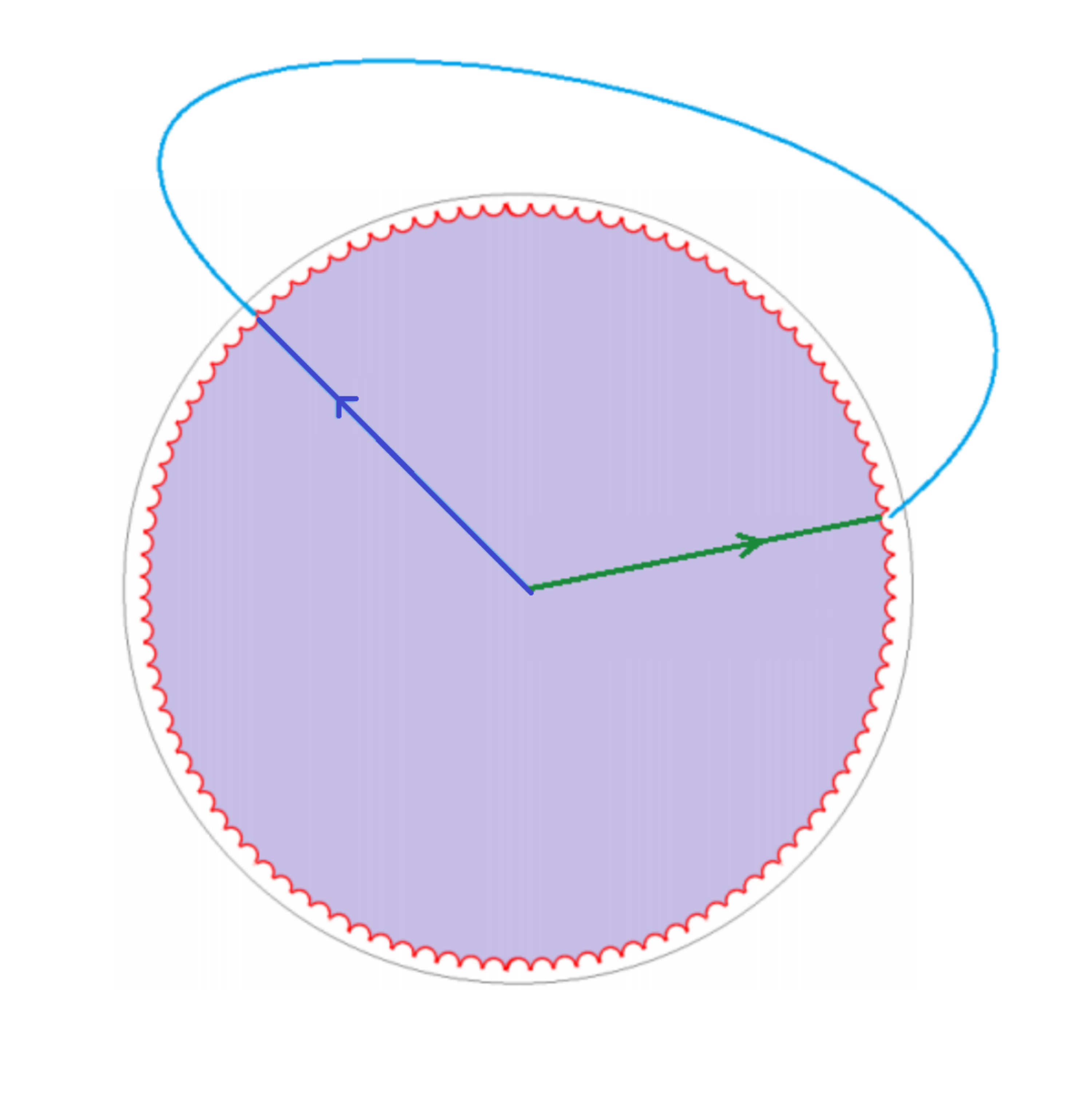}
\caption{The cut locus occurs at the polygon. Once the blue geodesic passes the cut locus
the shortest geodesic discontinuously jumps to the green line.}
\label{disc5}
\end{center}
\end{figure}

Let's come back to figure \ref{disc2} and follow the $shortest$ geodesic as the point $e^{-iHt}$ moves.  During the initial period the shortest geodesic is the portion of the straight blue line from the origin to the moving point. But something happens when the trajectory reaches the polygon  and the jump occurs---namely the cut locus is crossed. This is depicted in figure \ref{disc5}.

At the cut locus the blue geodesic is suddenly replaced by the green geodesic which, as soon as one moves a bit more, becomes the shorter of the two. There is not a discontinuous jump in the length of the geodesic, but there is a jump in the first derivative.

The parameter $t$ in $\aa(t)$ parameterizes the blue curve  in figure \ref{disc5} throughout its eternal history. Let's also parameterize the green curve, i.e., the minimal geodesic,  with a similar parameter $t'.$  
One may ask whether in the full high dimensional complexity geometry the green curve can be generated as a simple flow of the form  $e^{-iH't'},$ with a time-independent Hamiltonian $H'.$ Generically there is no reason to expect  that to be the case. In section \ref{subsec: Geodesics in Complexity Geometry} I explained the fact that most geodesics are generated by time-dependent Hamiltonians.  At the initial starting point $U=I$, the direction of the green geodesic will correspond to some linear combination of the generators $\sigma_I$---call it $H'(0).$ If $H'(0)$ is a mixed weight operator the trajectory that it generates by exponentiation  will not be a 
 geodesic of the full right-invariant complexity metric. However it will always be possible to represent it in terms of a unique time-dependent Hamiltonian $H'(t')$,   
\be
 U'(t')=Te^{-i\int H'(t')dt'}.
 \label{false}
 \ee
where $T$ means time-ordered.

  In what follows we assume that $H'(t')$  is approximately \kl, possibly with decreasingly small  contributions from increasingly  higher weight operators. 
  We will also assume that $H'(t')$  varies on a time scale which is not too short, so that the trajectory it generates is smooth,  but with inhomogeneities  along its length.

Let us imagine for a moment that the black hole actually evolved by the evolution indicated in   \ref{false}. As long as $H'$ is \kl \ we can expect that the wormhole that would be produces would have a classical geometry, but it would not be homogeneous. A reasonable guess is that it would correspond to a wormhole with matter distributed inhomogeneously along its length.

\begin{figure}[H]
\begin{center}
\includegraphics[scale=.5]{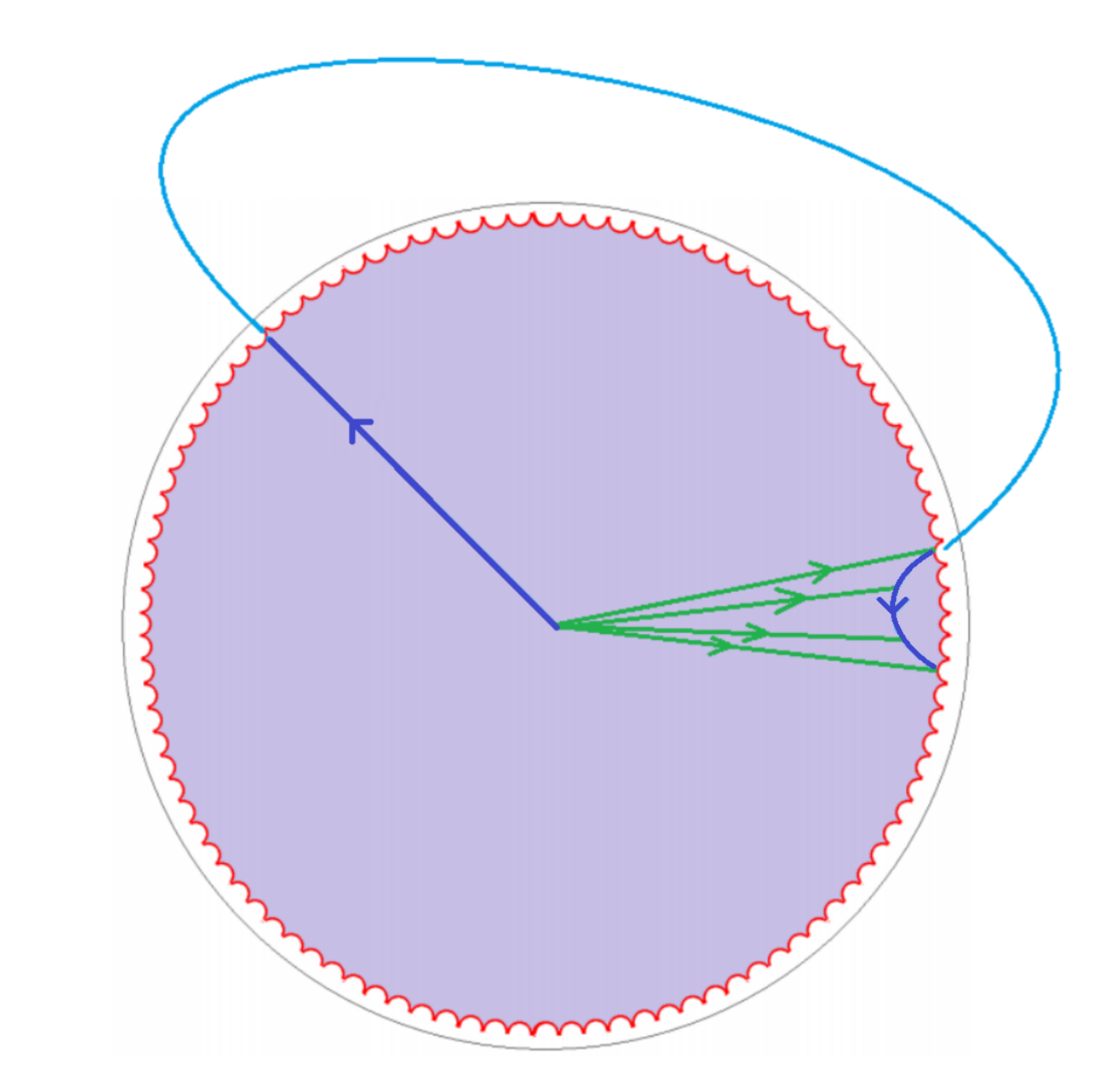}
\caption{Shortest geodesics beyond the cut locus.}
\label{disc3}
\end{center}
\end{figure}

   The fact that the blue and green geodesics wind up at the same place at the cut point can be expressed by the equation,
   \be
   e^{-iHt_c} = Te^{-i\int_0^{t_c} dt' H'(t')}.
   \ee
Suppose we continue the blue geodesic a bit further as in        
         figure \ref{disc3}.
We see that as we move along the small arc the (time-dependent) Hamiltonian that generates the shortest geodesic keeps changing continuously. Let's blow up the detail of the picture and then discuss it.

\begin{figure}[H]
\begin{center}
\includegraphics[scale=.6]{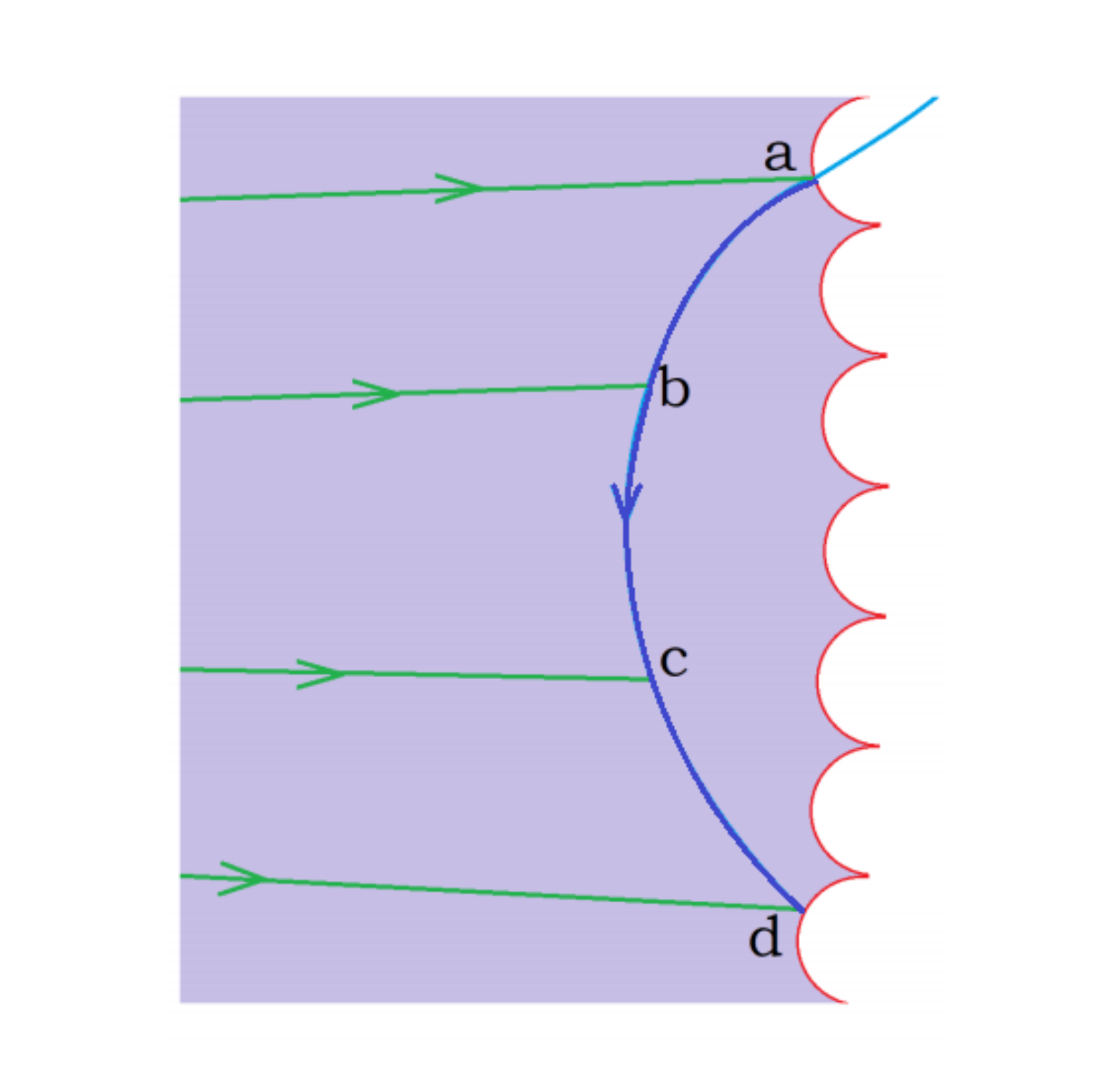}
\caption{Blowup of figure \ref{disc3}}.
\label{detail}
\end{center}
\end{figure}

\bn
The green geodesic  connects the identity element with the point $a$ in figure \ref{detail}. It has the form
   \be
 Te^{-i\int^a dt' H_a(t')}
   \ee

A little later the blue geodesic arrives at $b$. The new green geodesic has the form,

 \be
 Te^{-i\int^b  dt' H_b(t')}
   \ee

   Let us suppose for a moment that a black hole actually evolved by means of the time-dependent Hamiltonian $H_b(t')$. Assuming it is  
 approximately \kl, there is no reason why the wormhole it generates would  not be smooth and geometrical. It would be the wormhole we would expect for an evolving black hole in a smoothly time dependent background but one that was fine-tuned to reach the point $\bf{b}$.  In the present case all instantaneous expectation values at time $t_b$ would be consistent with that wormhole. Whether or not it is the $best \ way$  to describe the wormhole it would be $a \ way$  to describe it.
   
Let's go further and consider the  green geodesics that arrive at $c$ and $d.$  They have forms,
   \bea 
    & Te^{-i\int^c  dt' H_c(t')} \cr \cr
    & Te^{-i\int^d dt' H_d(t')}
   \eea
   while the  blue  geodesic  always has the form,
 $$ \aa(t) =e^{-iHt}.  $$


The mimimal distance (complexity) from the origin to $\aa(t)$ is continuous, as is the length of the wormhole, but the structure of the wormhole suddenly  changes at the cut point. In terms of tensor networks  the TN switches from a circuit generated by the true Hamiltonian $H, $ to one generated by $H_a(t').$   

Note that between $a$ and $b$  the complexity decreases a bit. According to CV so does length of the wormhole.
At some later  point between $b$ and $c$  the trajectory starts to move outward so that the complexity begins to increase. The continuous evolution proceeds until the point $d$ where it again exits at the polygon.
At that point another discontinuous jump  takes place. 
 
These jumps between the small arcs, and the increase and decrease of complexity between them, account for the small but  jagged fluctuations on the plateau  in figure \ref{C-evolution}. In terms of wormhole length the jumps imply a fluctuating, but overall constant wormhole length once $t_{exp}$ has been reached.

Every once in a (doubly exponential) while the trajectory comes out at just the right angle to find its way back to the origin. An explicit calculation shows that the probability for that is $e^{-2^N}$. Then we get a full quantum recurrence. In order for the trajectory to get all the way to the origin it must come in along the same line as it went out (as in figure \ref{disc4}). During the full recurrence the Hamiltonian would be the original time-independent $H$. The wormhole would be homogeneous with no matter,  but would shrink in length as one would expect from figure \ref{Whiteblack}, and then expand.  Partial recurrences in which the wormhole shrinks to some length $L$ which is large but much smaller than $2^N$ are not only possible, but also are far more frequent than full recurrences.

\section{Firewalls at Exp-time?}
It is an interesting question whether the inhomogeneities  of the wormhole would be detectable by someone crossing the horizon. In \cite{Susskind:2015toa} I raised the question of the possible existence of firewalls after \et. I still don't know the answer  in general,  but  there are two special cases where the answer is clear: the late time behavior of the exact TFD state, and certain doubly exponential times for a slightly perturbed TFD.

\subsection{The Exact TFD}\label{sub: Exact TFD}
Let's suppose that Alice  controls the left black hole and Bob jumps into the right black hole. The state at $t=0$ is the exact TFD. In the first case Alice does nothing and Bob waits an exponential time before jumping in. There is an exact boost symmetry generated by $(H_R-H_L)$  which can be used to  relate the problem to another case, one  in which Bob jumps in at time $t=0.$  Since we don't expect a firewall at $t=0$ we should not expect one at any jump-time. More precisely, the probability for Bob to detect a firewall, or any other matter behind the horizon, is independent of when he jumps in. If we assume that that probability is extremely small at $t_R=0,$ it will be equally small at any time.

\subsection{The Perturbed TFD}\label{sub: Perturbed TFD}
\begin{figure}[H]
\begin{center}
\includegraphics[scale=.3]{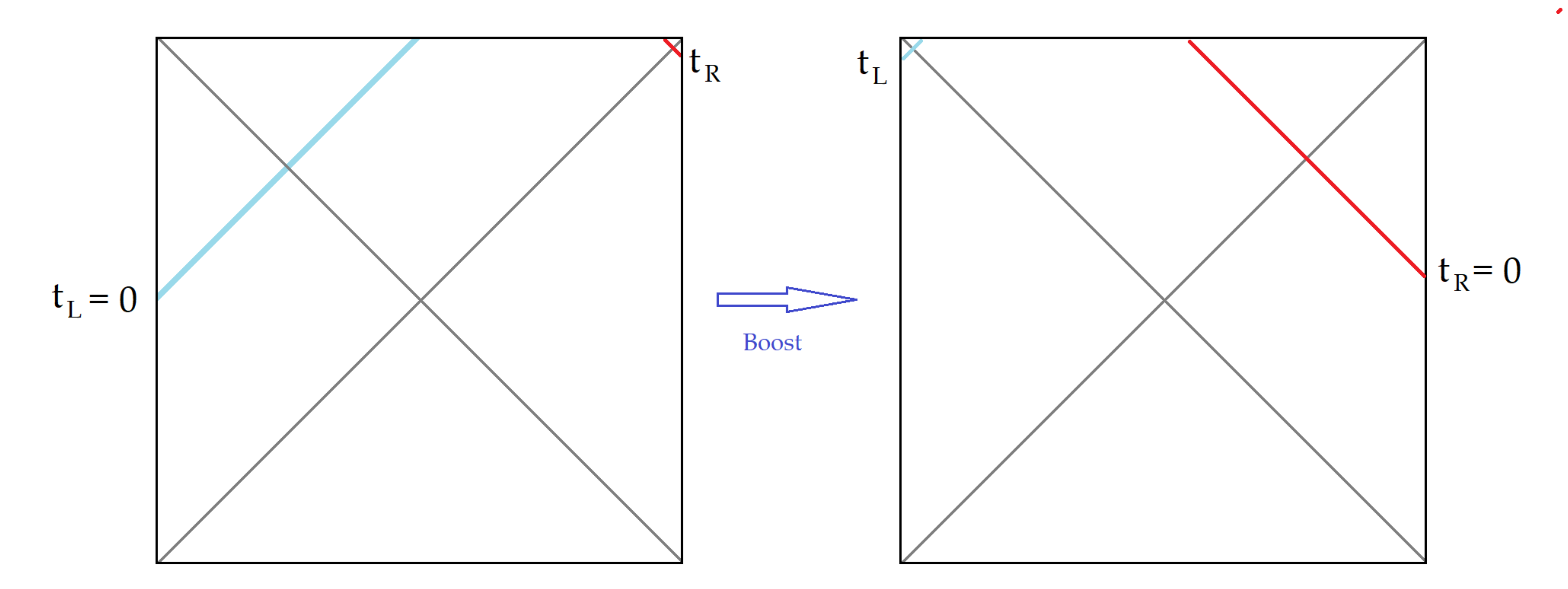}
\caption{Penrose diagram for Bob jumping into the right-side black hole when Alice has disturbed the left side. In the left panel the process is shown in a frame in which Alice disturbs her black hole at $t_L=0,$ and Bob jumps in much later at $t_R.$  The right panel shows the same process in a frame in which Bob jumps in at $t+R=0,$ and Alice perturbs her black hole at a late time.}
\label{AliceBob}
\end{center}
\end{figure}

In the second  case Alice perturbs the TFD state by applying a low energy perturbation at $t_L=0.$ The energy of the perturbation could be thermal or even lower as long as it is not exponentially lower. Bob jumps in at $t_R.$ The effect of Alice's perturbation is to break the boost symmetry so that Bob's subsequent experiences may not be $t_R$-independent.  Nevertheless, naive inspection of figure \ref{AliceBob} would suggest that the later Bob jumps in, the less  effect Alice's action will have  on him.
This reasoning is correct for sub-exponential $t_R$ but it is not necessarily correct for longer times. To see this we can first boost the diagram so that Bob always jumps in at $t_R=0,$ and Alice perturbs her black hole at a late time $t_L.$ Now, by an argument similar to the one about clocks in section \ref{sub: clocks}, the states generated by  perturbing at different $t_L$ are  approximately orthogonal to one another for $t_L < t_{exp}.$ Once $t_L > t_{exp} $ the quantum state becomes equivalent to a superposition of all those states for which the perturbation acted earlier. That includes states in which the perturbation acted in the past, and potentially may have created a shock wave that Bob would experience. 

While this argument  suggests that the naive argument is incorrect, it is not a definitive argument implying firewalls,  or even observable effects for time $t_L > t_{exp} $. 

However, there is one situation which is  clear. Suppose Bob waits \eet, until a short time $T$ before a full  quantum recurrence. Then  Bob will experience  whatever he would have experienced, had Alice made her perturbation at a small negative $t_L = -T.$
If $T$ is less than the scrambling time $t_*,$ then Bob will see a low energy quantum behind the horizon just before he and the quantum hit the singularity. If $T$ is of order a few scrambling times then Bob will be met by a very high energy
 shock wave \cite{Shenker:2013pqa}, i.e., a firewall.    If $T>>t_*$ the shock wave will be super Planckian and one might think this means that it will be even more destructive, but I don't know any calculation that would confirm this. For the moment, the question of what happens to Bob if he falls into the perturbed TFD at Exp-times is unanswered; there is no evidence that he experiences a firewall, but there is also no evidence that he doesn't. 

\section{Three  Descriptions: Which is Right?}

We can summarize the previous sections by three alternative hypotheses about what happens to wormholes at Exp-time and beyond. 
Let me list them and then comment.

\begin{enumerate}
\item  The ``You just keep going" theory: 

The wormhole just keeps growing indefinitely. This is  essentially the pseudo-complexity idea of \cite{Bouland:2019pvu}: the wormhole volume reflects the length of the geodesic $\aa(t)$ generated by the Hamiltonian $H,$  not the shortest geodesic\footnote{Pseudo-complexity as defined in  \cite{Bouland:2019pvu} allows a certain degree of shortening of a circuit by applying  local cancellations. However, as noted by the authors, these cancellations do not prevent the pseudo-complexity from growing forever. The authors themselves reject pseudo-complexity as a viable dual for wormhole volume.

  The term pseudo-complexity has also been used in very different context in  arXiv:2005.13801.}. Nothing happens at the cut locus; the bound on complexity does not limit the growth of wormholes.

According to this hypothesis the length a wormhole is proportional to the time measured by an ideal clock external to the CFT from when the black hole was created.

\item Long wormholes are quantum superpositions of shorter wormholes: 
As I discussed earlier the states of an evolving chaotic system, in time-steps of the Aharonov-Anandan time,  are approximately  orthonormal. After a time of order $2^N$ the system runs out of new states and enters into a quantum superposition of the earlier states, with approximately equal probability. From the boundary point of view this is correct, but the question is: what does  it imply about the geometry of the interior of a black hole? In particular does it mean that the state of the wormhole becomes a quantum superposition of vastly different classical geometries, and  that there is no way to describe it by a single geometry? 

The tensor network  model is illuminating here. At any given time the state of an evolving entangled system can be represented by a two-sided tensor network  as in figure \ref{TN}. 
The network is a rough model of the geometry of the wormhole \cite{Hartman:2013qma}\cite{Roberts:2014isa}.
\begin{figure}[H]
\begin{center}
\includegraphics[scale=.35]{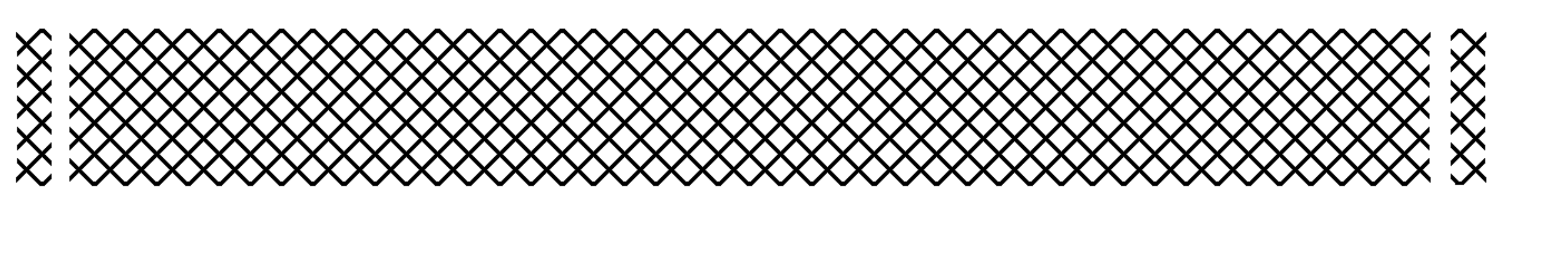}
\caption{A tensor network evolving by the addition of depth one layers on both sides. }
\label{TN}
\end{center}
\end{figure}
In each time-step the network is updated by adding a layer of gates to each side. As long as the tensor  network is not too long one can expect that it will be the shortest network of its type that can prepare the two-sided state.

But eventually when the length of the network exceeds $2^N,$ then just as with the Hamiltonian evolution of a clock, the new states cannot be linearly independent of all the previous states. The super-exponential tensor networks will describe states that are linear superpositions of sub-exponential tensor network states.

But that is not the only description of these states. By a counting argument similar to the argument for quantum circuits,  any state  can  be reached (to within a specified tolerance) by a \it single \rm  tensor network of no greater length than $2^N$. The local structure of the network will differ from the original, but it will nevertheless define a single geometry---not a superposition of geometries. This is the tensor network analog of what happens when the geodesic $\aa(t)$ reaches the cut locus.

This leads to the third alternative.

\item  Minimal geodesics in complexity geometry determine wormhole properties, and these minimal geodesics are never longer than the diameter of the geometry. As time evolves beyond Exp-time 
the wormhole geometries  make a series of closely spaced jumps,   smoothly
evolving between jumps, but the volume is continuous in time. These jumps are just the cut points described in  section \ref{sec: simple model}. They correspond to the fluctuations of complexity  on the complexity plateau.

\end{enumerate} 

 I have listed three possibilities so let me now comment on them in turn.

\bn

\begin{enumerate}
\item  The ``You just keep going" theory: 

This theory fails to give a reasonably compact description of the quantum state after Exp-time. We can see the problem most clearly by going to an extreme situation, in which a full quantum recurrence happens at some definite Expexp-time. The quantum state at that point, to within arbitrarily small error,  will have returned to the original TFD state---a state with vanishing wormhole length.
Left-right field correlations (of fields just outside the left and right horizons),  which we expect to decrease with increasing wormhole length, will be large at the recurrence time, indicating a short wormhole.  If Alice and Bob were to jump into their respective black holes shortly before the recurrence, they will meet in the wormhole.   Similarly, traversable wormhole experiments, which can only succeed if the wormhole is short, can succeed. In all ways the two-sided system will behave as if the wormhole length is very small or vanishing.

By contrast  the ``You just keep going" theory would imply  that the worm hole is of  doubly exponential  length, although  all observations on the holographic boundary theory, including those intended to probe the interior, will be consistent with a short wormhole. 

\item Long wormholes are quantum superpositions of shorter wormholes: 

From the  viewpoint of the holographic description of the black hole, this \it superposition \rm option is mathematically correct.  Any exterior correlations including ones which attempt to probe the interior will be correctly given by the quantum superposition of shorter wormholes. The problem is that we don't know how to interpret the superposition in terms of the properties of the interior. The presumed  boundary-bulk dictionary for the region behind horizons has two features which make the translation  of the boundary state extraordinarily difficult. The first is that the dictionary is extremely complex  \cite{Bouland:2019pvu}\cite{Susskind:2020kti};
 in fact it is exponentially complex, even for states of modest polynomial complexity\footnote{If that sounds contradictory it's not. The complexity of the dictionary for a given state is not the same as the complexity of the state. For example, to distinguish a state of modest complexity from a maximally complex state is typically exponentially complex. In \cite{Susskind:2020kti} I referred to this as the ``complexity of complexity."}. The dictionary for states of high complexity such as those at Exp-time is \it doubly  \rm exponentially complex.  

The other feature is that the dictionary is non-linear \cite{Papadodimas:2015jra}. Superpositions of states of the boundary theory generally do not map to superpositions of bulk states. 

\item The minimal geodesic (or minimal circuit, or minimal tensor network) theory determines the properties  of wormholes:

At any instant the full Hamiltonian evolution $U(t) =e^{-iHt}$ and the minimal geodesic arrive at  the same unitary operator, or the same two-sided entangled state. It follows that they both give the same results for all measurements on the boundary dual.  However, the minimal geodesic theory gives the most compressed version of a history that leads to that state. It says that all measurements at time $t$ are consistent with a wormhole history generated by 
the time-dependent  Hamiltonian $H'(t')$ evolving for a time less than or equal to $t_{exp}.$ While there is no jump in the volume or action of the wormhole, there are  jumps in the derivatives and in the detailed micro-structure of the wormhole.

These jumps are global in that the entire structure of the wormhole changes although the volume  is continuous. At the first cut point a uniform homogeneous wormhole will make a transition to an inhomogeneous state, and at later cut points  inhomogeneous wormholes transition to other inhomogeneous wormholes. 

\end{enumerate}

The three possibilities are all correct but they answer different questions. The first answers the question: How many effective  gates were used by the natural evolution of the black hole in order to arrive at the state at a given time? But as we've seen in the case of a quantum recurrence, this can be a grossly misleading estimate of the length.

The third answers the question of what is the most efficient way to get to that state, i.e., what is the smallest number of gates needed to arrive at the state? We could express it slightly differently: What is the least number of gates that it would take to shrink the wormhole back to zero length? In all cases this gives sensible results consistent with expectations about correlation functions, traversable wormhole experiments, and similar probes of the wormhole.

The second  possibility is about the quantum state of the system as described by the holographic dual boundary system and, because of the enormous complexity of the boundary-bulk dictionary, does not directly address the nature of the interior geometry.

\bn

The interesting fact is not that there are three different answers to three different questions for $t>t_{exp}$. What is remarkable is that for $t< t_{exp}$ all three  give the same answer; namely, there is a unique classical wormhole geometry whose length, volume, and action grow linearly with time in agreement with the classical Einstein equations. At exponential time the agreement breaks down and the three questions have different answers. 

\section{Conclusions}

Classical GR governs the interior of an eternal 
AdS black hole for a tiny fraction of the time\footnote{A fraction of order $\exp{(-\exp{S})}$}. The rare ``classical" episodes last for a time $\sim e^S$ and in between them   the black hole exists in a vast doubly exponential sea of time, stuck in a state of complexity equilibrium. 
Very little is known about the geometry of the interior, if indeed it has a geometry, during these periods of equilibrium. In this paper I've laid out what I know, which I will summarize  here:
 
 \bn
 
 \bi \item
The classical growth of the black hole interior shown in figure \ref{Whiteblack} cannot go on forever: the black hole  eventually runs out of linearly independent classical geometries, and a transition  must occur from the  complexity ramp to a complexity plateau. This plateau or equilibrium state is intermittently punctuated by Boltzmann-like complexity fluctuations, and if CV duality holds, fluctuations in the volume of the wormhole. A giant fluctuation---a.k.a. full quantum recurrence---appears as a bounce. It is only during such  fluctuations that the system follows the classical history described by the Penrose diagram in figure \ref{Whiteblack}. To put it concisely:

\bn

\it  The full Kruskal history  represented by figure \ref{Whiteblack} is a single Boltzmann-like complexity bounce. 

\rm
 
 \bn 
 
 \item
 Superpositions of very different  classical wormhole geometries can  be described as single geometries. One can understand this in terms of tensor networks. The state of the black hole can be represented by tensor networks in a number of ways.

 \begin{enumerate}
 \item 
The ``obvious" TN which consists of a number of elements which grows proportional to the age of the black hole. There is no limit to its size. For black holes older than \et \ it will be longer than  $e^S$.
 \item A linear superposition  of tensor network states, each shorter than $ e^S.$ This is analogous to a linear superposition of classically distinct geometries.
 \item A unique ``most efficient" TN. For old black holes the most efficient TN typically  has length  $\sim e^S$. 
 \end{enumerate}
 This suggests an equivalence between  wormholes longer than  $e^S$; quantum superpositions of sub-exponential wormholes; and wormholes of length  $e^S$ which represent maximally efficient circuits.
 
  For times shorter than  $e^S$ these three descriptions give the same classical geometry. The obvious tensor network and the most efficient tensor network are one and the same, and the superposition of geometries is trivial: it contains only one term. But beyond the TN analog of the cut locus the three descriptions diverge.
  
  \bn
 
 \item
 Complexity geometry provides a way to understand the entire history, and in principle determines a unique geometry. That geometry shrinks and grows during the Boltzmann bounce; it has constant exponential length with small fluctuations during complexity equilibrium; and exhibits  quantum recurrences on doubly exponential time scales. The  key geometrical concept for understanding the transition from complexity ramp to complexity plateau is the cut locus. Beyond the cut locus the evolution of the interior geometry evolves in a non-classical way, featuring a series of cut points that give rise to a fluctuating complexity at the top of the complexity plateau

\bn

\item
Once the age of the black hole exceeds $t_{exp}$  the shortest geodesic connecting the identity with $U(t) = e^{-iHt}$ is not the curve swept out by $e^{-iHt}.$ There is a shorter geodesic of the form $U'(t) =Te^{-i\int_{0}^a H(t') dt'    } $  where $H'(t')$ is an explicitly time-dependent Hamiltonian, not the actual Hamiltonian governing the evolution of the black hole.
However, the wormhole is indistinguishable from one that would have resulted from an evolution with a time-dependent $H'(t')$ for a time no larger than $t_{exp}.$ Such a wormhole will be inhomogeneous and of length  no longer than $e^S$. Two-sided experiments on the black hole, even of high complexity, will be consistent with such a wormhole. While the jump from $H$ to $H'(t)$ and from one $H'$ to another, is in-principle predictable, it is probably in practice random. These jumps in the structure of the wormhole are sudden, although the  length is continuous.

\bn

\item
I have  been unable to answer the question of whether, for the slightly perturbed TFD,  there are firewalls during complexity equilibrium.  The problem is closely related to another problem. If Alice perturbs her black hole at an early time, one or two scrambling times in the past, Bob if he jumps in at $t=0$ will encounter a Planck energy shock wave. But what happens if she perturbs her side in the very remote past? Nominally Bob will encounter an even more energetic shock wave.  But if we go far enough back to just the right doubly exponential time,  Bob will experience something very mild or nothing at all. While nominally the shockwave that he encounters has energy $\sim e^{e^S}$ the effect on Bob can be negligible. The naive idea that the higher the energy of a collision between Bob, and Alice's perturbation, the more damaging it will be, must  breakdown.

One might hope to understand super-high-energy collisions by studying them in flat space. We know what happens in flat space---a collision at super-high-energy creates a super-large black hole which would be far too big to fit into Alice's or Bob's black hole. How to think about exponentially high energy  collisions behind the horizon of a modest size black hole is not at all obvious.

\ei

\section*{Acknowledgements}

I am grateful to Adam Brown, Henry Lin, Michael Freedman, Phil Saad,   Steve Shenker, and Douglas Stanford    for discussions and insights about the material in this paper.

\end{document}